# CeMux: Maximizing the Accuracy of Stochastic Mux Adders and an Application to Filter Design

Timothy J. Baker

Department of Electrical Engineering and Computer Science, University of Michigan, bakertim@umich.edu

John P. Hayes

Department of Electrical Engineering and Computer Science, University of Michigan, jhayes@umich.edu

Stochastic computing (SC) is a low-cost computational paradigm that has promising applications in digital filter design, image processing and neural networks. Fundamental to these applications is the weighted addition operation which is most often implemented by multiplexer (mux) trees. Mux-based adders have very low area but typically require long bit-streams to reach practical accuracy thresholds when the number of summands is large. In this work, we first identify the main contributors to mux adder error. We then demonstrate with analysis and experiment that two new techniques, precise sampling and full correlation, can target and mitigate these error sources. Implementing these techniques in hardware leads to the design of CeMux (Correlation-enhanced Multiplexer), a stochastic mux adder that is significantly more accurate and uses much less area than traditional weighted adders. We compare CeMux to other SC and hybrid designs for an electrocardiogram filtering case study that employs a large digital filter. One major result is that CeMux is shown to be accurate even for large input sizes. CeMux's higher accuracy leads to a latency reduction of 4x – 16x over other designs. Further, CeMux uses about 35% less area than existing designs, and we demonstrate that a small amount of accuracy can be traded for a further 50% reduction in area. Finally, we compare CeMux to a conventional binary design and we show that CeMux can achieve a 50 – 73% area reduction for similar power and latency as the conventional design, but at a slightly higher level of error.

CCS CONCEPTS •Hardware~Integrated circuits~Logic circuits~Arithmetic and datapath circuits •Hardware~Emerging technologies~Analysis and design of emerging devices and systems• Hardware~Very large scale integration design~Application-specific VLSI designs~Application specific integrated circuits

**ACM Reference Format:**
To be automatically generated if accepted.

## 1  INTRODUCTION

Digital systems, like those in biomedical implants and the internet of things, [27] often have computational needs where approximate results are acceptable, but they must meet stringent implementation constraints such as low power and small size. Stochastic computing (SC) [3] has recently emerged as a promising computational paradigm for such applications. SC employs streams of random bits to process data, where the value of a bit-stream is related to the probability that its bits take value 1. A major advantage of this encoding scheme is that it enables the use of simple digital circuits to implement important arithmetic operations like multiplication and addition. SC is especially appealing for use in neural networks [17][18][19], image processing [8] or finite impulse response filters (FIRs) [4][5][6][7][10]. These applications rely heavily on the weighted addition operation which is expensive to implement in traditional non-stochastic or "binary" computing (BC). In contrast, weighted addition can be implemented very efficiently by SC using low-cost trees of multiplexers (muxes).

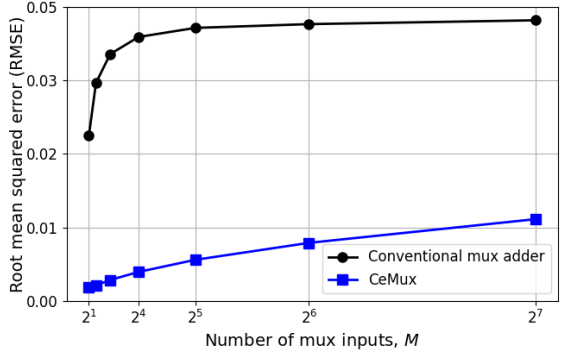

Figure 1: Root mean squared error vs. number of mux adder inputs *M* for a conventional mux-based SC adder (black) and CeMux (blue). SN length is $2^9$ bits.

Some prior studies of SC mux-based weighted adders for digital filters and similar applications [7][11][17] have found that muxes can have unacceptably low accuracy when the number of summands is large. This inaccuracy stems from SC's use of random bits which lead to random fluctuation errors in computed output values. Longer bit-streams mitigate the random errors, but they lead to unacceptable latency and high energy use. The inaccuracy of mux adders has increased interest in other, more expensive SC adders such as the accumulative parallel counter type [25].

Unexpectedly, recent work [1] has pointed out that small mux adders can be made far more accurate than expected via careful use of SC properties like correlation. That this can also be done efficiently for large, multi-input adders is a major conclusion of this paper. To illustrate, consider an experiment that compares a conventional mux adder [6][7] to our correlation-enhanced mux adder CeMux, which is presented in Sec. 4. Each adder is configured to implement weighted addition with $2^m$ summands and all weights randomly set to $\pm 1/2^m$ for a range of *m* values. In Fig. 1, both designs are simulated with input values that are randomly chosen from the interval $[-1,1]$. The estimated root mean squared error (RMSE) defined as

$$\text{RMSE}(Z, \hat{Z}) = \sqrt{\frac{1}{R}\sum_{i=1}^{R}(\mu_{Z_i} - \hat{\mu}_{Z_i})^2} \qquad (1)$$

is recorded, where $\mu_{Z_i}$ and $\hat{\mu}_{Z_i}$ are, respectively, the circuit's target value and actual output value during simulation run *i*. Fig. 1 shows that as *m* increases, CeMux's error is 3.4x to 12x lower than the conventional design's error.

In this work, we first identify the main error sources in mux adders through analysis that explores new angles of SC theory. This analysis leads to the concepts of precise sampling and full correlation which leverage aspects of correlation to reduce randomness in the operation of mux trees and thereby lower error levels. Implementing these techniques efficiently in hardware is at the heart of the CeMux design and the fact that it is orders of magnitude more accurate than alternative mux-based adders. The paper culminates in an application of CeMux to electrocardiogram (ECG) filtering.

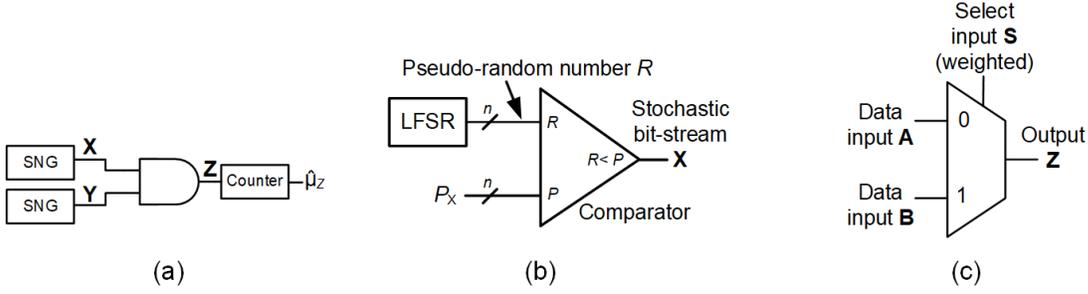

Figure 2: Stochastic computing circuits and elements. (a) Unipolar multiplier circuit; (b) Stochastic number generator based on an LFSR random number source and a comparator; (c) Multiplexer used in an SC weighted adder.

The key contributions of this work are:

1. Formulation of mux adder variance as the sum of three components which provides key insights into how variance can be decreased. Exact expressions are derived for mux adder variance in typical scenarios.
2. Introduction of the error-reduction techniques precise sampling and full correlation for mux adders.
3. Design of CeMux, a new mux adder design that exploits these techniques along with a low discrepancy number source to achieve high accuracy, even for large input sizes.
4. An ECG case study that demonstrates not only CeMux's superior accuracy, but also the fact that it greatly reduces both latency and area compared with existing SC designs.

## 2  BACKGROUND

First, we introduce relevant background information on stochastic computing. We adopt a slightly non-standard, but consistent notation for SC concepts that will simplify various equations.

### 2.1  Stochastic Computing Basics

#### 2.1.1  Stochastic Numbers

Stochastic computing (SC) uses (pseudo) random bit-streams called stochastic numbers (SNs) to encode and process information. An SN **X** is a stream of random bits $X_1 X_2 \ldots X_N$ which all have the same probability of taking value 1: $\mathbb{P}(X_i = 1) = P_X$ for $1 \leq i \leq N$. The numerical value $\mu_X$ of **X** is derived from $P_X$ and depends on the SN format used. Two popular formats are unipolar where $\mu_X = P_X$, and bipolar where $\mu_X = 2P_X - 1$. When using unipolar format, SN values are restricted to the unit interval [0,1] while bipolar format extends SN values to the negative domain $[-1,1]$. This work mainly uses the bipolar format, but the results also extend to the simpler unipolar case.

The usefulness of the stochastic encoding is best demonstrated with an example. Consider an AND gate with unipolar SN inputs **X**, **Y** and output **Z** (Fig. 2a). The output value of this simple circuit is $\mu_Z = \mathbb{P}(Z_i = 1) = \mathbb{P}(X_i \wedge Y_i = 1)$ which, if **X** and **Y** are uncorrelated, becomes $\mathbb{P}(X_i = 1)\mathbb{P}(Y_i = 1) = \mu_X \mu_Y$. Thus, $\mu_Z = \mu_X \mu_Y$ implying that the AND gate acts as a single-gate unipolar SN multiplier. Similarly, in bipolar format, an XNOR gate performs multiplication on uncorrelated inputs. The simplicity of multiplication in SC is a principal driving force for interest in the field.

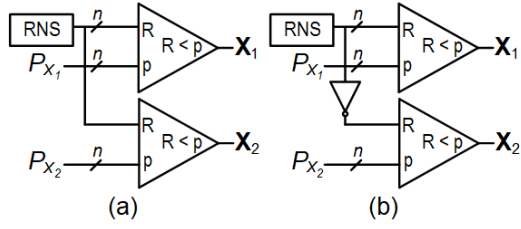

Figure 3: Generating cross-correlated SNs $X_1$, $X_2$. (a) maximally correlated SCC($X_1$, $X_2$) = 1; (b) maximally anti-correlated SCC($X_1$, $X_2$) = $-1$.

A drawback of the AND gate multiplier (or any single-output SC circuit) is that its output is not $\mu_Z$, but rather a stream of random bits $Z = Z_1 Z_2 \ldots Z_N$ which give an approximate estimate of $\mu_Z$. The most common estimator for unipolar SNs is the frequency of 1s in $Z$.

$$\hat{\mu}_Z = \frac{1}{N} \sum_{i=1}^{N} Z_i \qquad (2)$$

This estimator can be implemented with only a counter provided that the SN length $N$ is a power of 2 so that division by $N$ is effectively implemented by the radix point's implicit location in the counter's state. The estimator for bipolar SNs where $\mu_Z = 2P_Z - 1$ is similarly

$$\hat{\mu}_Z = 2\left(\frac{1}{N} \sum_{i=1}^{N} Z_i\right) - 1 \qquad (3)$$

which can be implemented with an up-down counter [3]. The difference between $Z$'s value $\mu_Z$ and $Z$'s estimated value $\hat{\mu}_Z$ is the circuit error which fluctuates due to the randomness of the bits used to measure $\hat{\mu}_Z$. Typically, the expectation of $Z$'s estimated value is very close to $Z$'s actual value, $\mathbb{E}[\hat{\mu}_Z] \approx \mu_Z$. However, subtle changes in the circuit's design can greatly affect the variance of $\hat{\mu}_Z$ and the circuit's accuracy. As exemplified in Sec. 3, understanding the relationship between logic design and the statistics of $\hat{\mu}_Z$ is crucial to managing error in SC.

On the input side of a stochastic circuit lie the stochastic number generators (SNGs) used to generate input SNs. A typical SNG consisting of a linear feedback shift register (LFSR) [3] and a comparator is shown in Fig. 2b. The LFSR's state $R$ cycles through integers in the range $[1, 2^n - 1]$ in a pseudo random order and acts as a pseudo uniform random number source. Each clock cycle, $R$ is compared to $P_X$ to produce $X_i$ where $\mathbb{P}(X_i = 1) \approx P_X$. By changing the control input $P_X$, the SNG generates SNs with desired values.

### 2.1.2 Stochastic Cross Correlation

Stochastic computing elements are usually designed to operate on statistically independent or uncorrelated SNs, although some designs require correlated inputs. When the intended level of input correlation is not realized, the circuit can have a biased output which can lead to serious errors [14][22]. The correlation between bits of two SNs, $X_1$ and $X_2$, is usually quantified by the stochastic cross correlation (SCC) metric, which measures the expected overlap between the 1s in $\mathbf{X}_1$ and $\mathbf{X}_2$ [14]. SCC takes values in [–1,1] where $\text{SCC}(\mathbf{X}_1, \mathbf{X}_2) = 0$ implies that $\mathbf{X}_1$ and $\mathbf{X}_2$ are uncorrelated. An $\text{SCC}(\mathbf{X}_1, \mathbf{X}_2)$ of +1 (–1) implies that the 1s in $\mathbf{X}_1$ and $\mathbf{X}_2$ overlap the maximum (minimum) number of times as determined by $P_{X_1}$ and $P_{X_2}$. For example, **A** = 010110 ($P_A = 1/2$) and **B** = 010010 ($P_B = 1/3$) have an estimated SCC of +1 because their 1s overlap as much as possible based on $P_A$ and $P_B$, while **A** and **C** = 101011 ($P_C$ = 2/3), have an estimated SCC of $-1$ because their 1s overlap as infrequently as possible based on $P_A$ and $P_C$.

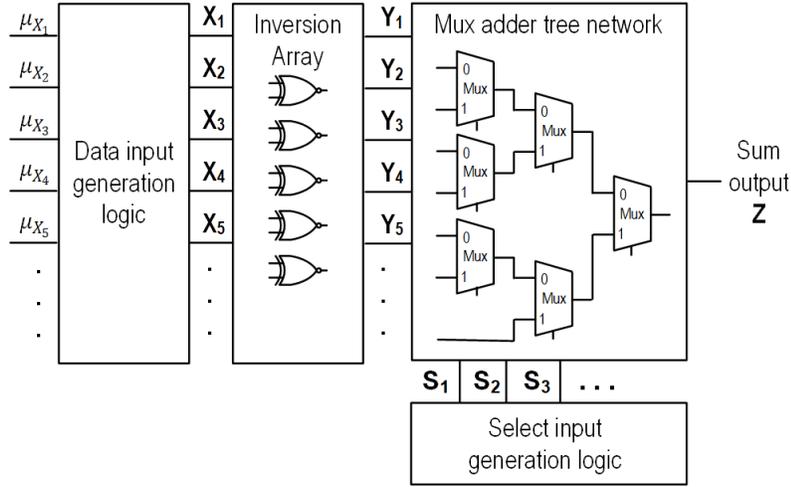

Figure 4: *M*-way weighted mux adder design. Each 2-way mux has a single select input; the input generation circuits may contain stochastic number generators in various configurations.

Two SNs, $\mathbf{X}_1$ and $\mathbf{X}_2$ can be generated with zero SCC by using separate and statistically uncorrelated RNSs in their SNGs. An SCC of +1 or –1 can be obtained by sharing an RNS between the SNGs, as shown in Fig. 3. Sharing in this manner saves hardware but increases the overall error if the circuit function requires uncorrelated inputs.

**2.2 Mux Adders**

We turn now to another fundamental SC operation, weighted addition, which is most often implemented by a multiplexer (mux), the focus of this paper. Fig. 2c shows a 2-input mux, which has two data inputs and a control input called the select line. If SNs **A** and **B** are applied to the data inputs, then the output **Z**'s value is a weighted sum of **A**'s value and **B**'s value where the sum's weights are determined by the value of an SN **S** applied to the select line. The weights of **A** and **B** must sum to one which enables a small range of adder types to be implemented and ensures that the sum is restricted to the probability range [0,1]. For instance, the mux in Fig. 2c is usually configured to compute

$$\mu_Z = \frac{1}{2}\mu_A + \frac{1}{2}\mu_B \qquad (4)$$

by setting **S**'s value $\mu_S$ to 1/2 implying that **A** and **B** are equally weighted. One way to understand (4) is to envision the mux as a sampling unit, where each clock cycle the control input **S** determines which input, **A** or **B**, is sampled, and has its bit propagated to the output. Since $\mu_S = 1/2$, **A** and **B** have an equal chance of being sampled each clock cycle implying that, on average, half of **Z**'s bits will be from **A** and half from **B**. Thus, **Z**'s value is given by (4). The viewpoint of muxes as sampling units will serve an important role in understanding how correlation can be used to improve the accuracy of mux adders.

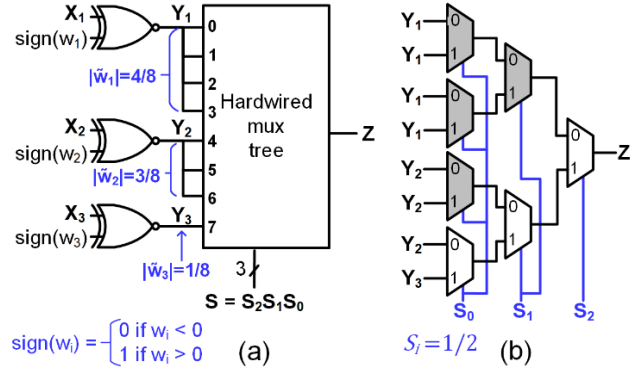

Figure 5: Three-input weighted mux adder that implements (6) and uses a hardwired mux tree [7]. The magnitudes of the normalized weights $\widetilde{w}_i = w_i / \sum_i w_i$ are encoded in the wiring of the data inputs of the mux tree and the signs of the weights are accounted for by the XNOR gate array.

The 2-input mux generalizes to an $M$-input mux circuit which computes

$$\mu_Z = \frac{1}{\sum_{i=1}^{M}|w_i|}\sum_{i=1}^{M} w_i \mu_{X_i} \qquad (5)$$

where $\mathbf{X}_1, \mathbf{X}_2, …, \mathbf{X}_M$ are bipolar input SNs with weights $w_1, w_2, …, w_M$, respectively. Conventional SC mux adders that compute (5) often have two stages: an XNOR gate array of bipolar multipliers followed by a mux adder tree as shown in Fig. 4a [4][6][7][10]. The XNORs multiply each bipolar data input with the sign of its corresponding weight. Then, each $\mathbf{Y}_i$ ($\mu_{Y_i} = \text{sign}(w_i)\mu_{X_i}$) is routed into a mux tree which computes $\mu_Z = \frac{1}{\sum_{i=1}^{M}|w_i|}\sum_{i=1}^{M}|w_i|\mu_{Y_i}$.

There are various mux tree designs that, together with an XNOR array, implement (5) and they mainly differ in how they use the mux select inputs to encode the normalized magnitude of each weight $|\widetilde{w}_i| = |w_i|/\sum_{i=1}^{M}|w_i|$. One basic design, the so-called "hardwired" mux tree [7] is best explained with an example. The hardwired mux tree in Fig. 5b computes

$$\mu_Z = \frac{1}{2}\mu_{Y_1} + \frac{3}{8}\mu_{Y_2} + \frac{1}{8}\mu_{Y_3}. \qquad (6)$$

Here, the mux select inputs $\mathbf{S}_2$, $\mathbf{S}_1$ and $\mathbf{S}_0$ all have value 1/2 and are shared amongst muxes on the same level of a full mux tree. With this configuration, all 8 mux tree inputs have probability 1/8 of being sampled each clock cycle. $\mathbf{Y}_2$ is then hardwired to three of the eight mux tree inputs because $|\widetilde{w}_2| = 3/8$. Likewise, $\mathbf{Y}_1$ is hardwired to half the mux tree inputs since $|\widetilde{w}_1| = 4/8$ and $\mathbf{Y}_3$ is hardwired to just one input because $|\widetilde{w}_1| = 1/8$. Thus, through hardwiring each $\mathbf{Y}_i$ to one or more input slots of the mux tree, weighted addition is implemented. In general, the height of the mux tree determines the values to which the normalized weights must be quantized, and Algorithm 1 describes the quantization procedure. In Fig. 5, the height is 3 and all $\widetilde{w}_i$ are quantized to 3-bit precision. The hardwired mux tree is most useful in resource-limited applications where the weights are not expected to be updated, such as in hearing aid filters [31] or electrocardiogram filtering [32].

---
Algorithm 1: Mux Tree Weight Normalization and Quantization
---

**Input:** weights $\mathbf{w} = [w_1, ..., w_M]$ and height of hardwired mux tree $m$
**Output:** absolute values of the quantized, normalized weights $[|\widetilde{w}_1|, ..., |\widetilde{w}_M|]$

$\mathbf{a}$ = *elementwise_absolute_value*($\mathbf{w}$)
$\mathbf{t}$ = $2^m \mathbf{a}/sum(\mathbf{a})$  // $\mathbf{t}$ is the numerator of the normalized weights which have denominator $2^m$.
$\mathbf{q}$ = *elementwise_round_to_nearest_integer*($\mathbf{t}$) // $\mathbf{q}$ is the quantized version of $\mathbf{t}$

// Sometimes after rounding, the quantized normalized weights do not sum to 1 and slight adjustments are
// are needed. $\mathbf{q}$ represents the numerator of these weights which have denominator $2^m$.

// If the sum of $\mathbf{q}$ exceeds $2^m$, decrement the numerator that results in smallest bias.
while $sum(\mathbf{q}) > 2^m$ do
   $i = \text{argmax}(\mathbf{q} - \mathbf{t})$
   $q_i = q_i - 1$
end

// If the sum of $\mathbf{q}$ is below $2^m$, increment the numerator that results in smallest bias.
while $sum(\mathbf{q}) < 2^m$ do
   $i = \text{argmax}(\mathbf{t} - \mathbf{q})$
   $q_i = q_i + 1$
end
**return** $\mathbf{q}/2^m$

---

In cases where weights are expected to be updated, the "biased selector" mux tree introduced in [5] is another design that can be used to implement weighted addition. In this case, the weights are not hardwired, but rather are encoded into the select input SNs' values which are no longer all set to be 0.5. When a change in summand weights is needed, the select input values can be updated. The weight flexibility comes at a high area cost, however, since many additional SNGs are needed for the select input SNs. Recent work aims at reducing this SNG overhead [4][6]; see Sec. 3 in [4] for a detailed explanation of the biased selector mux tree design. Note that the terminology "biased selector tree" is not used in [4], but was introduced here to help differentiate mux tree designs.

The central idea of this work is a new way to use correlation in mux trees. In [4][6], the authors use correlation to reduce the biased selector tree's SNG *area* while sacrificing as little accuracy as possible. Our work is distinct from [4][6] because here we use correlation to improve the *accuracy* of mux trees while incidentally decreasing their area. Additionally, our correlation techniques' impact on accuracy is validated by simulation in Sec. 5 as well as the analysis presented in Sec. 3 and the Appendix.

## 3 MUX ADDER ANALYSIS AND OPTIMIZATION

This section introduces our main accuracy enhancement techniques, precise sampling and full correlation, and demonstrates their effectiveness through a new analysis of mux tree errors.

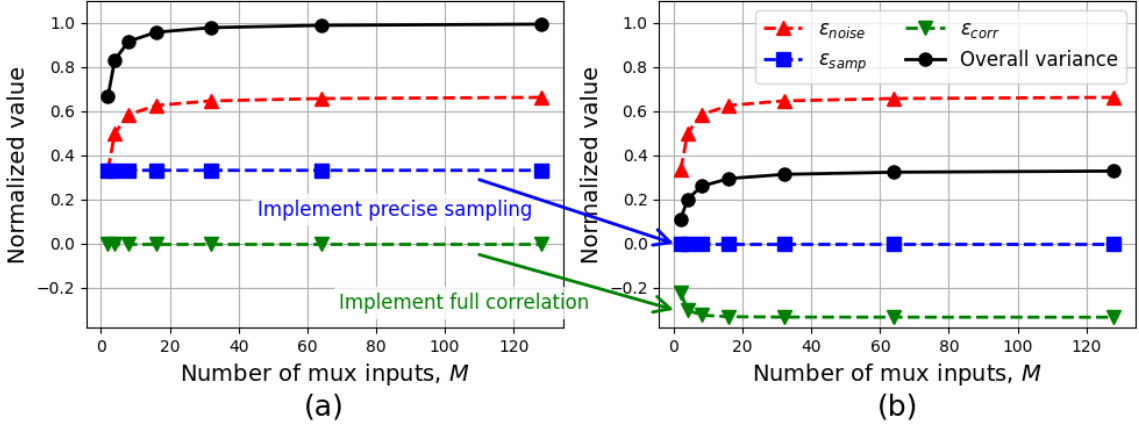

Figure 6: Variance of bipolar mux adders versus the number of mux inputs $M$ when input values are uniformly random and weights are randomly set to $\pm 1/M$. (a) An unoptimized mux circuit; (b) A mux circuit with precise sampling and full correlation implemented. All values are normalized by multiplying by $N$, the bitstream length.

### 3.1 Mux Adder Error

A stochastic circuit's error $\epsilon_Z$ is the difference between the output **Z**'s estimated value, $\hat{\mu}_Z$, found using a counter, and its target value, $\mu_Z^*$ found, in the case of mux adders, using (5):

$$\epsilon_Z = \hat{\mu}_Z - \mu_Z^* \tag{7}$$

The mean squared error, $\text{MSE}(\hat{\mu}_Z, \mu_Z^*) = \mathbb{E}[\epsilon_Z^2]$, can be expressed in the form of a bias-variance decomposition [2]

$$\text{MSE}(\hat{\mu}_Z, \mu_Z^*) = \text{Bias}(\hat{\mu}_Z, \mu_Z^*)^2 + \text{Var}(\hat{\mu}_Z) \tag{8}$$

where the bias (variance) quantifies the systematic (random) error of the circuit.

$$\text{Bias}(\hat{\mu}_Z, \mu_Z^*) = \mathbb{E}[\hat{\mu}] - \mu_Z^* \tag{9}$$

$$\text{Var}(\hat{\mu}_Z) = \mathbb{E}[(\hat{\mu}_Z - \mathbb{E}[\hat{\mu}_Z])^2] = \mathbb{E}[\hat{\mu}_Z^2] - \mathbb{E}[\hat{\mu}_Z]^2 \tag{10}$$

In the context of mux trees, the bias is error resulting from quantizing the input values and weights in (5) while the variance is error resulting from random fluctuations in the realized SN bit patterns. The variance is normally much greater than the bias in SC, so the bias is often approximated as zero. A key conclusion of the traditional Bernoulli model of SC is that the variance of any combinational circuit with bipolar output SN **Z** can expressed as

$$\text{Var}(\hat{\mu}_Z) = \frac{1 - \mathbb{E}[\hat{\mu}_Z]^2}{N} \tag{11}$$

where N is the length of **Z** [23]. Eq. (11) is useful because $\mathbb{E}[\hat{\mu}_Z]$ is easy to compute for a given circuit. For instance, the variance of a mux adder that computes (5) can be expressed as

$$\text{Var}(\hat{\mu}_Z) = \frac{1 - \left(\sum_{i=1}^{M} \widetilde{w}_i \mu_{X_i}\right)^2}{N} \tag{12}$$

where $\widetilde{w}_i = w_i/\sum_{i=1}^M |w_i|$ and $\mu_{X_i}$ have been quantized to the precision of the circuit.

In many cases, (11) and (12) overestimate the variance of a circuit that employs typical LFSR SNGs (Fig. 2b) because these SNGs do not produce Bernoulli-type SNs. Instead, the hypergeometric SN model can be used to obtain better estimates of variance for LFSR SNGs [1].

We now present a new formula for mux circuit variance that flexibly applies to both the Bernoulli and hypergeometric SN models:

$$\text{Var}(\hat{\mu}_Z) = \epsilon_{noise} + \epsilon_{samp} + \epsilon_{corr} \tag{13}$$

Formal definitions and derivations for $\epsilon_{noise}$, $\epsilon_{samp}$, and $\epsilon_{corr}$ are given in the Appendix. Here, we focus on a high-level explanation of (13).

Like (12), $\epsilon_{noise}$, $\epsilon_{samp}$ and $\epsilon_{corr}$ all depend on the SN length N, the input values, and the weights. Decomposing mux variance into these three specific components highlights how precise sampling and full correlation improve mux accuracy. For instance, Fig. 6 illustrates how $\epsilon_{noise}$, $\epsilon_{samp}$ and $\epsilon_{corr}$ vary with the number of mux inputs M for a mux adder that has random bipolar input values and all weights randomly set to $\pm 1/M$. Fig. 6a corresponds to a conventional mux adder and shows that as the number of inputs increases, the variance quickly saturates to a high value of $1/N$ [11]. Fig. 6b corresponds to a mux adder with our proposed precise sampling and full correlation methods. It shows that precise sampling reduces $\epsilon_{samp}$ to zero, and full correlation pushes $\epsilon_{corr}$ below zero to about $-1/(3N)$. Together these techniques lead to a significant overall variance reduction of 67%. In Sec. 5, we will show that this accuracy improvement is further amplified when a low discrepancy RNS is used in place of an LFSR-type RNS.

Similar to (13), other equations that decompose stochastic circuit error into distinct components have been proposed in research that does not specifically focus on mux adders. In [16], the authors express circuit error as a sum of approximation, quantization, and fluctuation errors. The former two errors are systematic and together constitute the bias of the circuit whereas the latter error is the variance. In [22], the authors express error with a correlation term and a variance term. Their correlation error is a non-negative bias term and is distinct from $\epsilon_{corr}$ which can take negative values and quantifies how correlation affects circuit variance. In [24], the authors introduce a hypergeometric decomposition method which maps a circuit's variance into a function of the input values and input variances. Our analysis differs from most prior error analyses [7][11][16][22][23] because, like [1] and [24], it applies to the hypergeometric model of SNs which much better represent SNs derived from typical LFSR SNGs.

### 3.2 Sampling Error and Precise Sampling

Mux addition relies on sampling the input SNs, and $\epsilon_{samp}$ measures the variation of that sampling process. For instance, consider the mux tree of Fig. 5b with independent inputs $\mathbf{Y}_1$, $\mathbf{Y}_2$, $\mathbf{Y}_3$ that have length $N = 16$. Since $|\widetilde{w}_2| = 3/8$, $\mathbf{Y}_2$ is expected to be sampled six times ($|\widetilde{w}_2|N = 6$), but $\mathbf{Y}_2$ may be sampled anywhere from zero to sixteen times due to random fluctuations in the mux select inputs $\mathbf{S}_2$, $\mathbf{S}_1$ and $\mathbf{S}_0$. When an input is not sampled its expected number of times, the mux output will be biased thus causing error that is characterized by $\epsilon_{samp}$. If $C_i$ is the number of times bipolar input $\mathbf{Y}_i$ is sampled, then for the mux tree in Fig. 5b,

$$\epsilon_{samp} = \frac{1}{16^2} \sum_{i=1}^{3} \sum_{j=1}^{3} \mu_{Y_i} \mu_{Y_j} \text{Cov}(C_i, C_j) \tag{14}$$

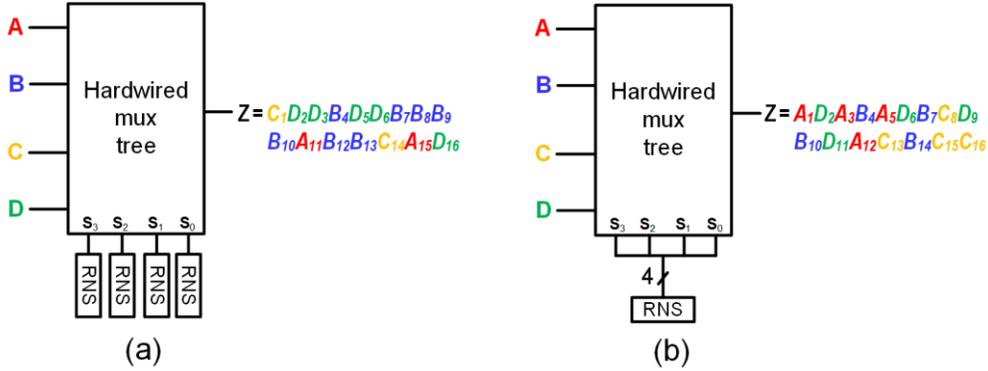

Figure 7: Sampling inputs from a hardwired mux tree where each input is equally weighted. (a) Noisy sampling where the select lines are generated from independent RNSs; (b) precise sampling where the select lines are derived from a single RNS state. Note how in (b) each evenly weighted input is sampled exactly 25% of the time.

where $\text{Cov}(C_i, C_j)$ is the covariance between $C_i$ and $C_j$ and $\text{Cov}(C_i, C_i) = \text{Var}(C_i)$. Eq. (14) highlights the fact that $\epsilon_{samp}$ (and thus part of the mux's variance) is dependent on the variation in the number of times each input is chosen. Reducing this variation motivates the concept of precise sampling.

Let $N$-bit SNs $\mathbf{Y}_1, \mathbf{Y}_2, \ldots, \mathbf{Y}_M$ with normalized weights $|\widetilde{w}_i|$, be input to a mux tree and let $C_i$ be the number of times $\mathbf{Y}_i$ is sampled by the tree. A mux tree performs *precise sampling* when, with probability 1, each input is sampled its expected number of times, up to a rounding error. Formally, precise sampling is when $\mathbb{P}(|C_i - \mathbb{E}[C_i]| < 1) = 1$ for $1 \leq i \leq M$. After quantizing $\mathbb{E}[C_i] = |\widetilde{w}_i|N$ to the nearest integer, implementing precise sampling guarantees that $\epsilon_{samp}$ is always zero.

Conventional hardwired mux trees such as HWA [7] do not perform precise sampling because separate and independent RNSs are used to feed the mux select lines, and fluctuations between these RNSs causes sampling variation as seen in Fig. 7a. Instead, when the SN length is $2^n$, precise sampling can be performed by a height $h$ hardwired mux by deriving the mux select lines from the state of a *single* RNS as shown in Fig. 7b. A key feature of the chosen RNS is that it generates numbers from $[0, 2^n - 1]$ without repetition. This construction ensures that each mux input slot is sampled exactly $2^{n-h}$ times and eliminates variation in the number of times an input is sampled implying that $\epsilon_{samp} = 0$. Moreover, our new construction (Fig. 7b) replaces the $h$ RNSs of the conventional design (Fig. 7a) with a single RNS thus saving considerable area. Suitable choices for a precise sampling (pseudo) RNS include an $n$-bit LFSR with the all-0 inserted to its state sequence making it nonlinear [1], and an $n$-bit counter.

### 3.3 Full Correlation

Fig. 3 shows two SNGs sharing a single RNS. The sharing of a single RNS amongst the $M$ data inputs of a mux adder is common practice in SC because it saves considerable area and because correlation among the mux data inputs was believed to have no effect on output error. This assumption of correlation insensitivity was disproven recently when it was shown that correlation can greatly increase the accuracy of a mux [1].

To illustrate, consider the mux tree in Fig. 8a with four uncorrelated inputs $\mathbf{Y}_1, \mathbf{Y}_2, \mathbf{Y}_3, \mathbf{Y}_4$ each with unipolar value 1/2, length 8 and weight 1/4. During each clock cycle $i$, some of the input bits $(Y_{1,i}, Y_{2,i}, Y_{3,i}, Y_{4,i})$ are 0 and others 1. By happenstance, it is possible for the mux to propagate a 0 every single clock cycle resulting in $\mathbf{Z} = 00000000$. In this case, $\mathbf{Z}$'s estimated value is 0 which poorly represents $\mathbf{Z}$'s actual value of 1/2.

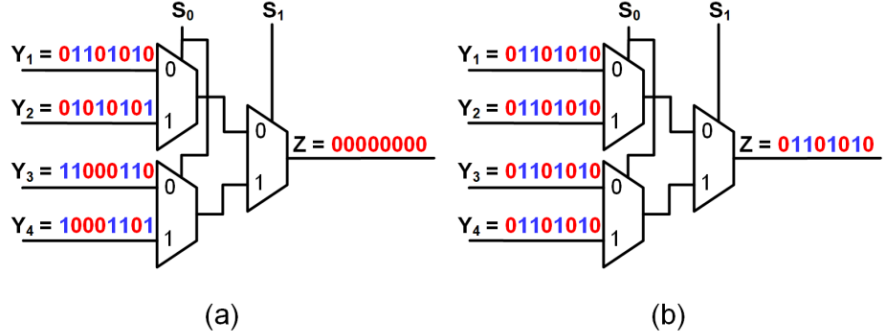

Figure 8: Effect of data correlation on mux behavior; (a) Possible outcome of mux tree with uncorrelated inputs; (b) unique outcome of mux tree with maximally correlated inputs of the same value.

In contrast, Fig. 8b shows the same mux adder configuration, but when $Y_1$, $Y_2$, $Y_3$, $Y_4$ are maximally correlated by sharing an RNS. Here, $Y_1$, $Y_2$, $Y_3$, $Y_4$ are identical since they all have the same value and share an RNS. Regardless of the mux select inputs, the output **Z** will always be a copy of one of the mux's four identical inputs and **Z**'s estimated value is the correct value of 1/2.

In general, when the data inputs have different values, correlation amongst mux tree data inputs lessens the possibility of high error caused by the mux selection process (as in Fig. 8a). Correlation's impact on accuracy (quantified by $\epsilon_{corr}$ in (13)) is greatest when correlation is maximized. Thus, to maximize accuracy, mux trees should achieve *full correlation* which occurs when the SCC is +1 between all pairings of mux data inputs. Full correlation can be achieved by careful sharing of RNSs; an example is given in our CeMux design (Sec. 4.1) whose mux tree achieves full correlation. Note that full correlation can be applied to improve accuracy when using a hardwired mux tree and also when using a biased selector mux tree like the designs in [4][6].

## 4  CEMUX

Now we formally introduce CeMux, a bipolar weighted mux adder which combines our two correlation-inspired techniques, precise sampling and full correlation, with other recent advances in SC to form a particularly efficient design. CeMux (Fig. 9) implements weighted addition (5) using an XNOR multiplier array and hardwired mux tree as in the general mux adder structure of Fig. 4. Since the weights are fixed and known ahead of time, Fig. 9 simplifies the XNOR array of Figs. 4 and 5a by explicitly showing that inputs with negative weights are inverted by the XNOR array and inputs with positive weights are unmodified by the XNOR array. The following subsections explain the remainder of the design in a component-wise manner, while Algorithm 2 summarizes the overall design procedure.

### 4.1  Data Input RNS and Full Correlation

Like other mux adders [4][6][7], CeMux uses a single RNS to generate its data input SNs. Recently, it has been shown that stochastic circuit's accuracy can be improved by using low discrepancy (LD) sequence generators as the RNS for SN generation [9][19][28][29]. LD sequences are deterministic sequences sometimes used to emulate a sequence of uniformly random numbers, but with lower variance. Well-known examples are the Halton, Sobol and van der Corput sequences [28][29]. Examples of a LD-generated SNs (LD SNs) are **A** = 10101010 ($P_A = 1/2$) and

## Algorithm 2: Designing a CeMux Adder

**Input:** weights $w = [w_1, \dots, w_M]$, precision $n$ and SN length $2^n$

***Data RNS:*** Introduce an $n$-bit Sobol LDS generator [9]. Connect its output to the probability conversion (PCC) array.

***Correlation Inverters:*** Use a set of $n$ inverters to generate the inverted Sobol RNS value which is also connected to the PCC array.

***PCC Array:*** Construct an array of $n$-bit comparators that compare data inputs $\mu^+_2, \mu^+_2, \dots \mu^+_{X_k}$ with positive weights to Sobol RNS value and compare inputs, $\mu^-_{X_1}, \mu^-_{X_2} \dots \mu^-_{X_{M-k}}$ with negative weights to the inverted Sobol RNS value. The output of this array is a set of $M$ SNs, $\mathbf{X}_1, \mathbf{X}_2, \dots \mathbf{X}_M$.

***Sign Inverter Array:*** Place inverters on all $\mathbf{X}_i$ whose $w_i < 0$. Leave other $\mathbf{X}_i$'s untouched. The output of this array is a set of SNs $\mathbf{Y}_1 \dots \mathbf{Y}_M$. Note this inverter array is equivalent to the XNOR array of conventional mux designs [4][6][7].

***Precise Sampling RNS:*** Assemble an $n$-bit counter whose $i$-th MSB is connected to the select line of all muxes on the $i$-th level of the mux tree (the tree root is level 1).

***Hardwired Mux Tree:*** Utilize Algorithm 1 with inputs $w$ and $n$ to derive the absolute values of the normalized weights $|\widetilde{w}_1| \dots |\widetilde{w}_M|$. Each $\mathbf{Y}_i$ input is hardwired to $|\widetilde{w}_i| 2^n$ mux tree input slots. The output of this tree is CeMux's output SN **Z**.

***Output Counter:*** Insert an $n$-bit up-down counter that increments when **Z**'s bit is 1 and decrements when **Z**'s bit is 0. The output of this counter is **Z**'s estimated value $\hat{\mu}_Z$.

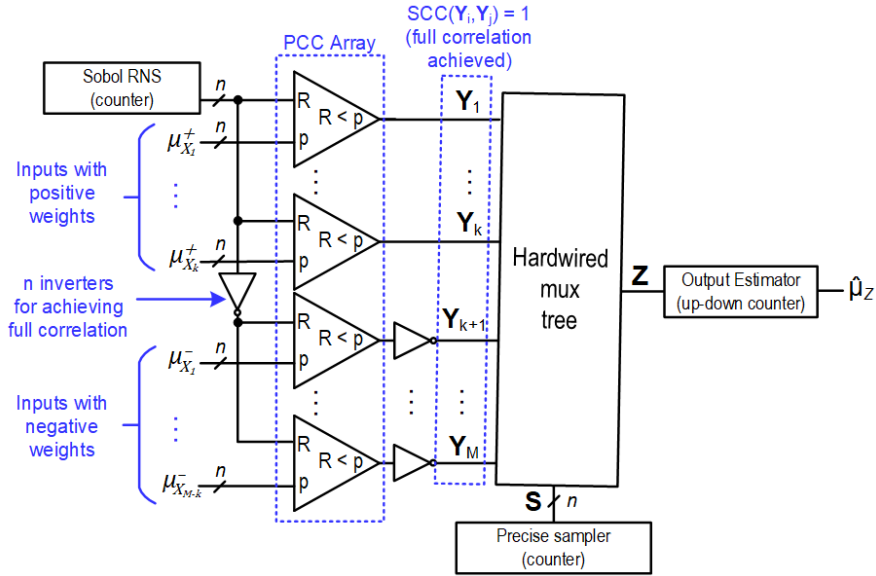

Figure 9: The proposed CeMux design for computing (5). Instead of an array of XNOR gates, inverters are explicitly shown for inputs with negative weights.

$\mathbf{B} = 11101110$ ($P_B = 3/4$) which exhibit the key feature of LD SNs—the 1s are roughly uniformly distributed throughout the SN rather than randomly distributed. This uniform distribution of 1s reduces random fluctuation in the circuit operation and can often lead to more accurate results. The accuracy improvement is more significant when the circuit has few RNSs such as in the CeMux design which uses just a single RNS. Thus, the simplest Sobol sequence generator (implemented using the reverse state of a standard counter [9]) serves as CeMux's RNS.

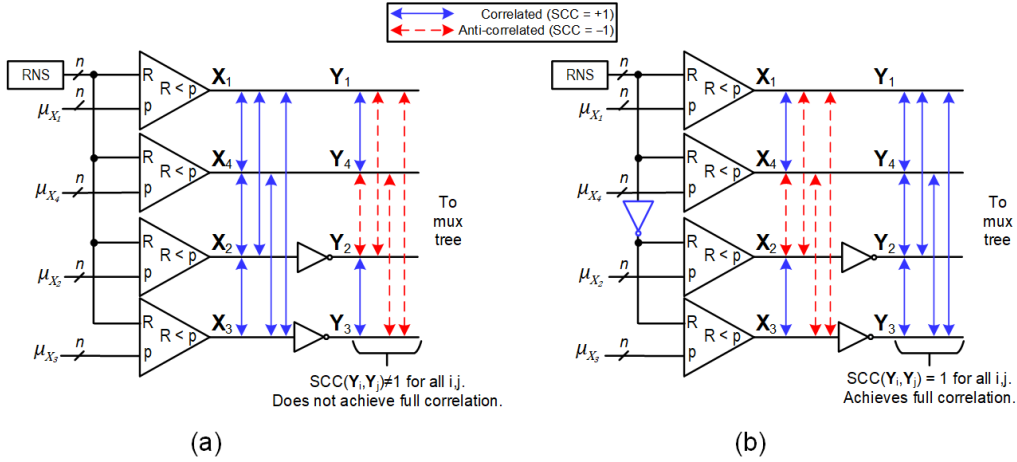

Figure 10: Visualizing the correlation of mux tree inputs **Y₁, Y₂, Y₃, Y₄** with $w_1, w_4 > 0$ and $w_2, w_3 < 0$. (a) conventional SN generation where all SNs are generated using the same RNS output. (b) full correlation generation method where SNs with positive weights are generated using the RNS output while SNs with negative weights are generated using the inverted RNS output. All data SNs in (b) are fully correlated as they enter the hardwired mux tree.

CeMux's mux tree achieves full correlation when $SCC(Y_i, Y_j) = +1$ for all *i,j*. Ensuring this happens is not as straightforward as simply sharing the RNS amongst all data input SNGs. Fig. 10a demonstrates this point where a mux containing four inputs **X₁, X₂, X₃, X₄** with corresponding weights $w_1, w_4 > 0$, and $w_2, w_3 < 0$ is shown. The pairwise SCC amongst **X₁, X₂, X₃, X₄** is 1. However, since $w_2, w_3 < 0$, **X₂** and **X₃** are inverted which results in $SCC(Y_1, Y_4) = SCC(Y_2, Y_3) = 1$ and $SCC = -1$ for all combinations of **Y₁, Y₂, Y₃, Y₄**. Thus, full correlation is not achieved for the mux tree in Fig. 10a.

In contrast, CeMux's proposed SNG configuration for these SNs is shown in Fig. 10b. A single RNS is shared by the SNGs, but the RNS output is inverted for **X₂**'s and **X₃**'s SNGs. The result is $SCC(X_1, X_4) = SCC(X_2, X_3) = 1$, and $SCC = -1$ for all other SN pairings. Following the inversion of **X₂** and **X₃**, we have that $SCC(Y_i, Y_j) = 1$ for all *i,j* and full correlation is achieved by the mux tree. More generally, to achieve full correlation in CeMux, all inputs share an RNS, but inputs with negative weights use the inverted RNS output for SN generation while inputs with positive weights use the unaltered RNS output for SN generation.

### 4.2 Probability Conversion Circuits

The comparator used in an SNG can be generalized to what is known as a probability conversion circuit (PCC) and another choice for an SNG's PCC is a weighted binary generator (WBG) [21]. A study by Zhong et al. [6] on mux adders showed that using WBGs in place of comparators can reduce circuit area by over 50% because about half of the WBG circuit can be shared amongst all SNGs. WBGs' area efficiency suggests that they may be useful in CeMux, however, extensive simulation experiments show that input SNs cannot be reliably correlated when WBGs replace comparators in SNGs. In other words, full correlation cannot be achieved with WBGs, implying that their use degrades CeMux's accuracy. To maximize accuracy, comparators are used as CeMux's PCCs. Nevertheless, due to the WBG's impressive area efficiency, we evaluate a version of CeMux that uses WBGs in the case study (Sec. 5.3).

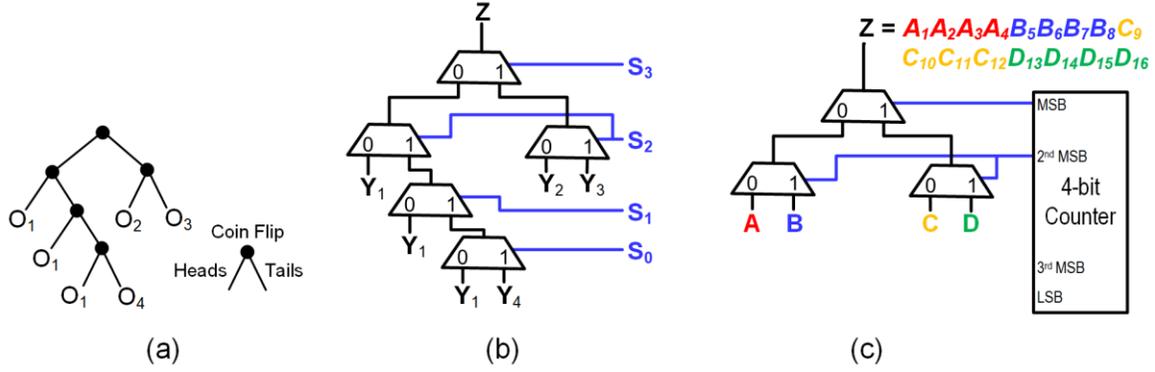

Figure 11: CeMux's hardwired mux tree implementation. (a) DDG tree corresponding to random variable $X$ with distribution $\mathbb{P}_X(O_1) = 7/16$, $\mathbb{P}_X(O_2) = \mathbb{P}_X(O_3) = 1/4$ and $\mathbb{P}_X(O_4) = 1/16$; (b) corresponding DDG-style hardwired mux tree that implements (15). (c) using a counter to implement precise sampling for a mux tree that implements $\mu_Z = 1/4(\mu_A + \mu_B + \mu_C + \mu_D)$.

## 4.3 Hardwired Mux Tree and Precise Sampling

CeMux's hardwired mux tree height is set to $n$ where $2^n$ is the SN length. This is the largest tree height that enables precise sampling to function fully and yields the lowest quantization error. Increasing $n$ might seem to imply a large increase in hardware due to the exponential increase in mux numbers. That is not the case, however, because many of the 2-way muxes in a hardwired mux tree have identical data inputs and thus can be eliminated. For instance, all the shaded muxes in Fig. 5b can be removed, reducing the mux count from seven to three. In general, the number of muxes in a hardwired mux tree grows linearly with $n$.

The scaling of the number of non-redundant muxes in a hardwired mux tree can be understood by relating it to Knuth and Yao's discrete distribution generating (DDG) trees for random number generation [15]. A DDG tree describes an algorithm that uses a series of fair coin flips to sample from the given discrete distribution. Each internal node in a DDG tree corresponds to a coin flip and each leaf node corresponds to an outcome of the given distribution; an example is show in Fig. 11a. In terms of our work, each 2-way mux in a hardwired mux tree corresponds to a DDG tree internal node, each hardwired mux input $Y_1, Y_2, ..., Y_M$ corresponds to a DDG tree leaf node and the normalized weights $|\widetilde{w}_1|, |\widetilde{w}_2|, ..., |\widetilde{w}_M|$ correspond to the DDG tree's target distribution. Bearing in mind this connection, Knuth and Yao's work tells us that an optimal height-$n$ hardwired mux tree can be constructed simply by inspecting the $n$-bit binary expansions of the normalized weights.

For example, consider the hardwired mux tree construction in Fig. 11b which computes

$$\mu_Z = \frac{7}{16}\mu_{Y_1} + \frac{1}{4}\mu_{Y_2} + \frac{1}{4}\mu_{Y_3} + \frac{1}{16}\mu_{Y_4} \qquad (15)$$

This construction can be arrived at in the following manner. Let level 1 be the root of the mux tree. First, $Y_1$ is connected to a mux on level 2, level 3, and level 4 of the tree because $\widetilde{w}_1$'s binary expansion ($0.0111_2$) has a 1 in the $2^{-2}$, $2^{-3}$ and $2^{-4}$'s place. Likewise, $\widetilde{w}_2 = \widetilde{w}_3 = 0.0100_2$ implies $Y_2$ and $Y_3$ should both be connected to a mux on level 2 of the tree. Finally, $\widetilde{w}_4 = 0.0001_2$ implies $Y_4$ is to be connected to a mux on level 4.

Generally, the DDG method of constructing a height-$n$ hardwired mux trees implies that the total number of muxes in the simplified tree is one less than the total number of 1s in the $n$-bit binary expansions of the normalized weights. For an $M$-input height-$n$ hardwired mux tree, the (often impossible) worst-case is when every weight has an all-1s binary expansion. Combining this worst-case with the consideration that a height-$n$ mux tree can have at most $2^n - 1$ muxes implies a (loose) upper bound of $\min(Mn - 1, 2^n - 1)$ on the maximum number of muxes. Hence, the number of muxes grows linearly rather than exponentially with $n$.

CeMux implements precise sampling by using an $n$-bit counter's state as the mux select input lines. The counter's $i$-th MSB is connected to muxes on the $i$-th level of CeMux's hardwired mux tree as in Fig. 11c. With this construction, the output **Z** tends to consist of runs of bits from the same SN (also seen in Fig. 11c). Since CeMux uses a low-discrepancy RNS that uniformly distributes the 1s in each input SN, this method of precise sampling leads to a highly accurate output.

## 5 EXPERIMENTAL EVALUATION

Finally, we evaluate the effectiveness of full correlation and precise sampling experimentally and compare CeMux to alternative weighted-adder designs.

### 5.1 Full Correlation and Precise Sampling with Random Data

We first consider using CeMux to compute (5) in the case where $w_i$ and $\mu_{X_i}$ are randomly chosen from $[-1,1]$. CeMux's precision $n$ is fixed at 10, and the number of inputs ($M$ in (5)) is varied from 8 to 256. The SN length is $2^{10}$ and $R = 5{,}000$ simulation runs are used. The RMSE (1) is measured when simulating CeMux both with and without full correlation and/or precise sampling. The results are shown in Fig. 12a. The RMSE for every configuration increases as $M$ is increased, which is consistent with other mux adder studies [4][6][7][11].

Full correlation is removed from CeMux by deleting the $n$ leftmost inverters in Fig. 9. Removing full correlation increases the RMSE by about 25% for all values of $M$, as shown in Fig. 12a. On the other hand, when precise sampling is removed from CeMux (by replacing the bottom RNS in Fig. 9 with a set of $n = 10$ LFSRs), the RMSE rises by about 114-630% depending on $M$. This larger RMSE increase indicates that precise sampling improves accuracy much more than full correlation. Finally, CeMux without either correlation technique has the worst RMSE which is about 160-810% higher than full CeMux's RMSE.

We then repeated the foregoing experiment but replaced CeMux's Sobol RNS with an LFSR to evaluate the usefulness of the low discrepancy RNS. The results are plotted in Fig. 12b and show that CeMux with the Sobol RNS is more accurate than CeMux with an LFSR RNS. Thus, the low discrepancy sequence complements our correlation techniques and further improves accuracy. For additional reference, also plotted in Fig. 12b is the RMSE when using HWA from [7]. In all, CeMux is found to be 1.5x – 5.7x more accurate than CeMux with an LFSR, and 2.6x – 9.3x more accurate than HWA, a standard mux adder design.

### 5.2 CeMux Variants

In the previous section, we observed that accuracy is degraded by removing full correlation or precise sampling from CeMux. These correlation techniques always maximize accuracy, but useful accuracy-area trade-offs are possible and may sometimes be worth considering. First, as discussed in Sec. 4.2, WBGs can be used in place of comparators to save a large amount of area in return for the loss of full correlation. Second, CeMux's precise sampling method is built around a hardwired mux tree whose design cannot readily adapt to a change in summand weights. A biased selector mux tree can replace CeMux's hardwired mux tree resulting in a design that implements full correlation, but not precise sampling. Importantly, this design variant can adapt to changes in summand weights by updating the appropriate registers.

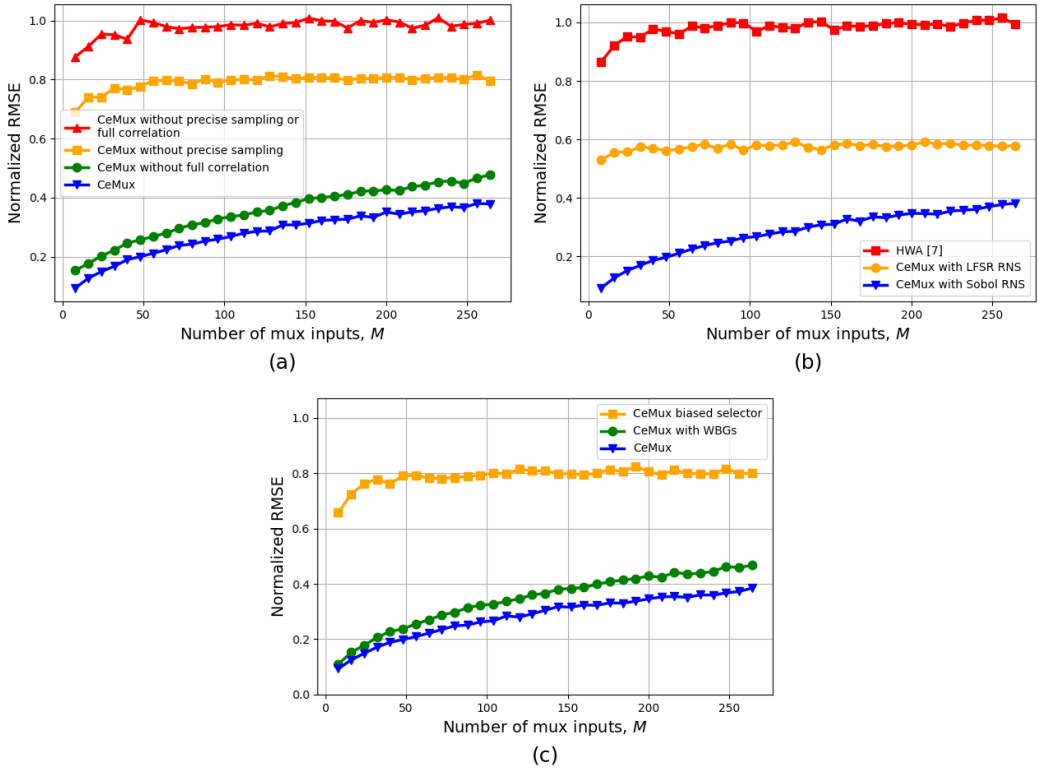

Figure 12: Errors vs. number of mux inputs for various mux adders given random bipolar input values and weights. SN length is $2^{10}$ and RMSE is normalized by multiplying by $\sqrt{2^{10}}$. (a) RMSE for CeMux with and without its correlation techniques; (b) RMSE for CeMux with and without its Sobol RNS, as well as RMSE for HWA; (c) RMSE for CeMux and two of its variants.

To explore these tradeoffs further, the following analysis considers CeMux along with two variations on its design. The first variant, "CeMux with WBGs" has better area than CeMux while the second variant "CeMux biased selector" has more flexibility in updating summand weights, but worse area. However, both designs will have worse accuracy than CeMux. Demonstrative of this point is Fig. 12c which uses the same experimental set-up as Sec. 5.1 but has these CeMux variants as the designs under consideration. The data in Fig. 12c reveals that the CeMux variants have lower accuracy than CeMux and the trends in the RMSE data match that of Fig. 12a where correlation techniques were explicitly removed from CeMux.

### 5.3 ECG Case Study

Next, we present an electrocardiogram (ECG) filtering case study which evaluates the performance of CeMux and other SC designs in the context of a practical application in the biomedical device field. Analysis of denoised ECG signals is used to monitor patient cardiovascular health and, for example, detect conditions like arrhythmia [12]. Noise is often removed from an ECG signal with a finite impulse response (FIR) filter, but FIR designs tend to place large computational demand on an ECG monitor's limited computational resources [26][32][32]. The simplicity of SC digital filters suggests a promising direction for filter design in the ECG domain. In this case study, we demonstrate that CeMux is the most promising SC design candidate for ECG digital filtering.

Table 1: SC Mux-based Filter Design Specifications and Features

| Design Name | Mux Tree Type | Mux Data PCC | Mux Select RNS | Mux Select PCCs | Full Correlation? | Precise Sampling? | Adaptable Weights? |
|---|---|---|---|---|---|---|---|
| CeMux* | hardwired | comparators | counter | --- | Yes | Yes | No |
| CeMux with WBGs* | hardwired | WBGs | counter | --- | No | Yes | No |
| CeMux biased selector* | biased selector | comparators | LFSRs | WBGs | Yes | No | Yes |
| Basic hardwired | hardwired | WBGs | LFSRs | --- | No | No | No |
| Basic biased selector | biased selector | WBGs | LFSRs | WBGs | No | No | Yes |

* Indicates our proposed design and its variants.

*5.3.1 Digital Filter Design*

An $M$-tap digital finite impulse response FIR filter implements $Z_i = \sum_{j=0}^{M-1} h_j X_{i-j}$ where the $\{X_i\}$ are samples from a digital signal, $\{h_i\}$ are the constant filter coefficients, and $\{Z_i\}$ is the filtered signal. Filters with more taps tend to perform better filtering, but at the cost of higher computational resources such as more multipliers. Eq. (5) is a scaled version of the filter equation and thus mux adders are well-suited for SC-based filter design [5].

Muscle contractions, device noise and electrosurgical noise are three major ECG noise types that can be modeled by random noise [20]. We thus add random noise to a benchmark ECG signal [12][13] to generate a suitable test input. Then, as is common practice in filter design, we utilize MATLAB to derive the coefficients of an $M$-tap lowpass filter with a cutoff frequency of $0.1\pi$ rad/sample. The purpose of this lowpass filter is to remove the high frequency noise from the ECG signal.

Next, we compare CeMux and its variants mentioned in Sec. 5.2 to other SC designs for the ECG filter application. One design we compare against is built around a hardwired mux tree (like HWA [7]) while another is built around a biased selector mux tree (like MWA [7] and the designs found in [4][6][10]). We also compare our designs with a typical accumulative parallel counter (APC) design [25]. For the APC case, an array of XNOR gates is used to perform SC multiplication between each signal input and its corresponding filter coefficient. Then, each product bit-stream is accumulated by the parallel counter which exhaustively counts all the incident bits. Only two Sobol RNSs [9] are needed in an $M$-input APC design, one RNS for the $M$ signal inputs and one RNS for the $M$ filter coefficients.

Table 1 summarizes the mux-based SC designs under consideration and their features. All designs employ a shared Sobol RNS for the mux data input SNs which gives the best accuracy. WBGs are used and shared [6] whenever possible because they need less area than comparators. Designs that feature a hardwired mux tree do not need PCCs for the mux select input SNs because those SNs always have value 0.5 and so can be produced from the RNS directly. We do not consider ad hoc accuracy-sacrificing techniques like circular shifting which degrade accuracy to improve area [4]. For all designs including the APC case, the bit-width of all RNSs, counters, WBGs and comparators is set to $n$ bits where $n$ is varied throughout the case study. Bit-stream length is always set to $N = 2^n$.

*5.3.2 Accuracy and Latency Analysis*

First, each hardware filter design listed in Sec 5.3.1 is simulated with the derived filter coefficients and the noisy ECG signal as input. The RMSE (1) is estimated using $R = 10{,}000$ simulation runs when the precision of each circuit is set to $n = 10$ bits and the number of filter taps $M$ is varied. The results are plotted in Fig. 13a. CeMux's RMSE ranges from 4x to 12x lower than all the other SC designs, indicating its superior accuracy. Further, CeMux maintains its low RMSE as $M$ is increased and never exceeds an RMSE of $2^{-7}$, indicating that CeMux remains accurate with large input sizes. CeMux's variants are also significantly more accurate than the traditional SC designs because they each implement one of our two correlation techniques.

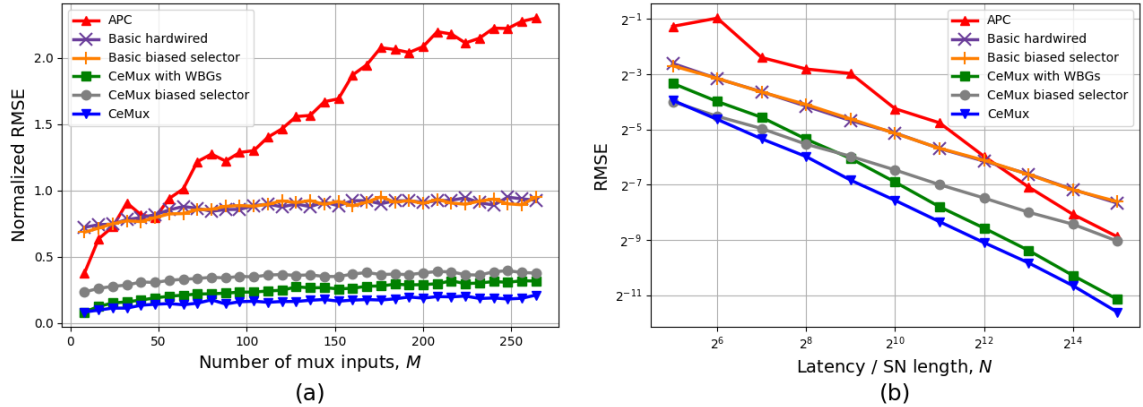

Figure 13: Error analysis of various SC digital filters for ECG case study. (a) RMSE versus number of inputs, $M$; (b) RMSE versus latency and SN length, $N$. RMSE is normalized in (a) by multiplying by $\sqrt{2^{10}}$.

Surprisingly, CeMux's accuracy far exceeds that of the APC design, which usually has very high accuracy due to the use of an expensive parallel counter to perform deterministic summation [19][25]. In the present case, however, we find RNS sharing leads to high summation error caused by highly correlated intermediate errors. Not sharing RNSs would improve accuracy but lead to unreasonable area cost. Note that RNS sharing does not always lead to high error in the APC design, which has been shown to perform accurately in neural networks [19].

Next, to understand the latency of each design, we perform a similar experiment where the input size $M$ is fixed to 150 while the precision $n$ and bit-stream length $N = 2^n$ are varied. Fig. 13b plots the RMSE estimated by simulation against the latency $N$ of each design. The latency needed for CeMux to surpass certain accuracy thresholds is always much lower than its SC counterparts. For instance, CeMux achieves an RMSE below $2^{-4}$ with 64-bit SNs whereas the other mux designs require 256-bit SNs to meet the same target. The slope of the CeMux curve is also steeper than the other mux designs' curves indicating CeMux's latency improvement increases when more stringent accuracy thresholds are required. Overall, to hit a given accuracy threshold, CeMux and its variants require a latency around 4x to 16x lower than their counterpart SC designs.

### 5.3.3 Area Analysis

Next, we use Synopsys Design Compiler with the Nangate 45nm open cell library to synthesize the various filters and estimate the area of their weighted addition datapath. We do not consider the memory used to store prior signal values since each design requires the same amount of memory. Fig. 14a shows CeMux's component-wise area breakdown for an $n = 10$-bit precision design that implements an $M = 100$ tap filter. CeMux's 100 comparators take up 80% of the overall area, indicating efforts to improve CeMux's area should target these components. One such approach would be to decrease the bit-width of CeMux's comparators which saves area but will increase the quantization error of CeMux. For instance, changing CeMux's comparator's bit-width from 10 to 9 bits would reduce the comparator area by about 10% and while increasing the quantization error from $\sim \frac{1}{2^{10}}$ to $\sim \frac{1}{2^9}$ [7].

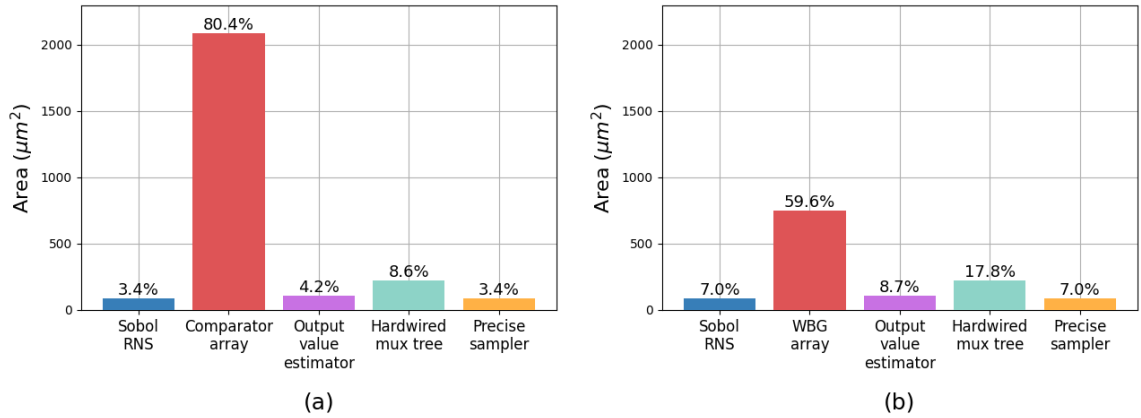

Figure 14: Component wise area breakdown for (a) CeMux; (b) CeMux with WBGs. Precision $n$ is 10 bits and the number of filter taps $M$ is 100.

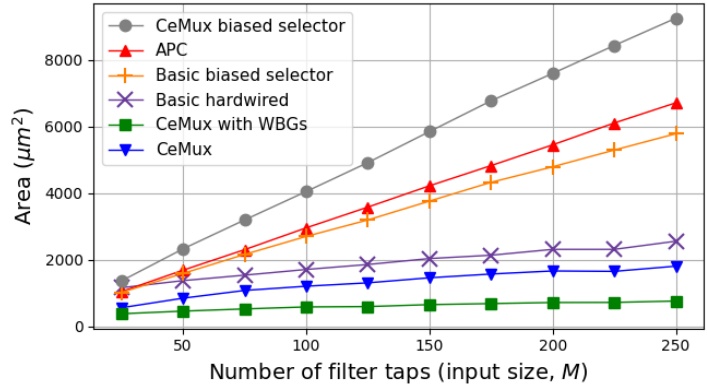

Figure 15: Area vs. input size for various SC digital filter designs. Precision $n$ is 10 bits for all designs.

For comparison, we also synthesized CeMux with WBGs and plotted the results in Fig. 14b. In this case, the WBGs take up 60% of the overall area. The total circuit area of CeMux with WBGs is 50% less than the standard CeMux design which, along with Fig. 13, again indicates that using WBGs can lead to area savings at the cost of accuracy.

Finally, we synthesize the various SC filter designs while varying the filter size. Fig. 15 plots the circuit area versus number of filter taps $M$ for $n = 10$ bit precision. CeMux is smaller than other SC designs, achieving an average 35% area reduction over other conventional SC designs because it replaces costly SNGs for weights or mux select inputs with a simple but precise sampling counter. As before in Fig. 14, Fig. 15 shows that using WBGs instead of comparators in CeMux reduces the area by about half due to the WBGs' area efficiency. Thus, at the expense of accuracy (see Fig. 13) the area of CeMux can be further reduced by employing WBGs.

Table 2: Cost and Performance of a CeMux Filter and a Sequential Binary (SB) Filter

| | 10-bit CeMux Filter | | | 8-bit Sequential Binary Filter | | |
|---|---|---|---|---|---|---|
| Input Size, $M$ | Area ($\mu m^2$) | Power ($\mu W$) | RMSE ($\times 10^{-3}$) | Area ($\mu m^2$) | Power ($\mu W$) | RMSE ($\times 10^{-3}$) |
| 25  | 566  | 13.16 | 4.17 | 1158 | 20.67 | 5.06 |
| 50  | 853  | 20.16 | 4.47 | 1385 | 23.45 | 3.49 |
| 75  | 1085 | 25.71 | 4.57 | 1603 | 28.51 | 3.99 |
| 100 | 1212 | 28.82 | 4.87 | 1761 | 29.64 | 3.61 |
| 125 | 1314 | 31.42 | 5.16 | 1855 | 30.83 | 3.13 |
| 150 | 1465 | 35.52 | 5.41 | 2066 | 36.58 | 4.69 |
| 175 | 1578 | 37.50 | 5.62 | 2119 | 37.99 | 4.85 |
| 200 | 1670 | 39.79 | 5.64 | 2293 | 39.71 | 3.55 |
| 225 | 1658 | 39.44 | 6.26 | 2324 | 39.70 | 4.19 |
| 250 | 1813 | 43.20 | 6.05 | 2477 | 41.70 | 4.12 |

Fig. 15 also shows that the three largest designs are conventional biased selector, APC, and CeMux biased selector. These designs have higher area because they use more SNGs and because, unlike the other designs, they are flexible in their ability to update filter coefficients stored in an external memory whose cost is not considered here. CeMux biased selector is the costliest design because it uses comparators rather than WBGs as the PCCs. Importantly, however, this CeMux variant's accuracy leads to better latency than the APC and the conventional biased selector designs (Fig. 12b) making it a suitable alternative or accuracy stringent applications. While flexibility in updating filter coefficients is a convenient feature, FIR designs for resource limited applications like ECG filtering [32] and hearing aids [31] often assume and benefit from fixed filter coefficients.

*5.3.4 Comparison with Binary Computing*

The focus of this work has been on the analysis and improvement of SC mux adders. We have demonstrated that CeMux is the best mux-based SC adder in terms of accuracy (Fig. 13) and area (Fig. 15). For completeness, we also give a brief comparison with a conventional binary design. We compare a 10-bit CeMux filter with a traditional sequential binary (SB) filter designed using MATLAB's Filter Design HDL coder. The SB design is synthesized assuming the filter coefficients are fixed, and the SB design employs standard optimizations like the exploitation of symmetric coefficients which greatly reduces multiplier count. Note that the designs' precision levels are chosen to give them similar accuracy. Both designs are also configured to operate in real-time which requires each one to process digitized ECG samples at the sampling rate of 360 Hz.

Table 2 shows the area, power and RMSE of the CeMux and SB filters as the filter length is varied from $M = 25$ to $M = 250$. CeMux's area is 49% to 73% lower than the SB design's area due to the use of cheap SC computational units. The CeMux design must process the entire 1024-bit SN at a rate of 360 Hz to meet the real-time latency constraint. Consequently, CeMux's digital clock frequency is set to be faster than the SB's design digital clock frequency which results in both designs having similar power despite CeMux having lower area. Finally, the SB design has better RMSE, especially as input size grows. Based on the data presented in Table 2, we conclude that CeMux, a mux-based SC adder, has the potential of being a lower-cost alternative to conventional binary designs. Besides being smaller, SC designs also offer greater fault tolerance [7] which is one avenue for future exploration with CeMux.

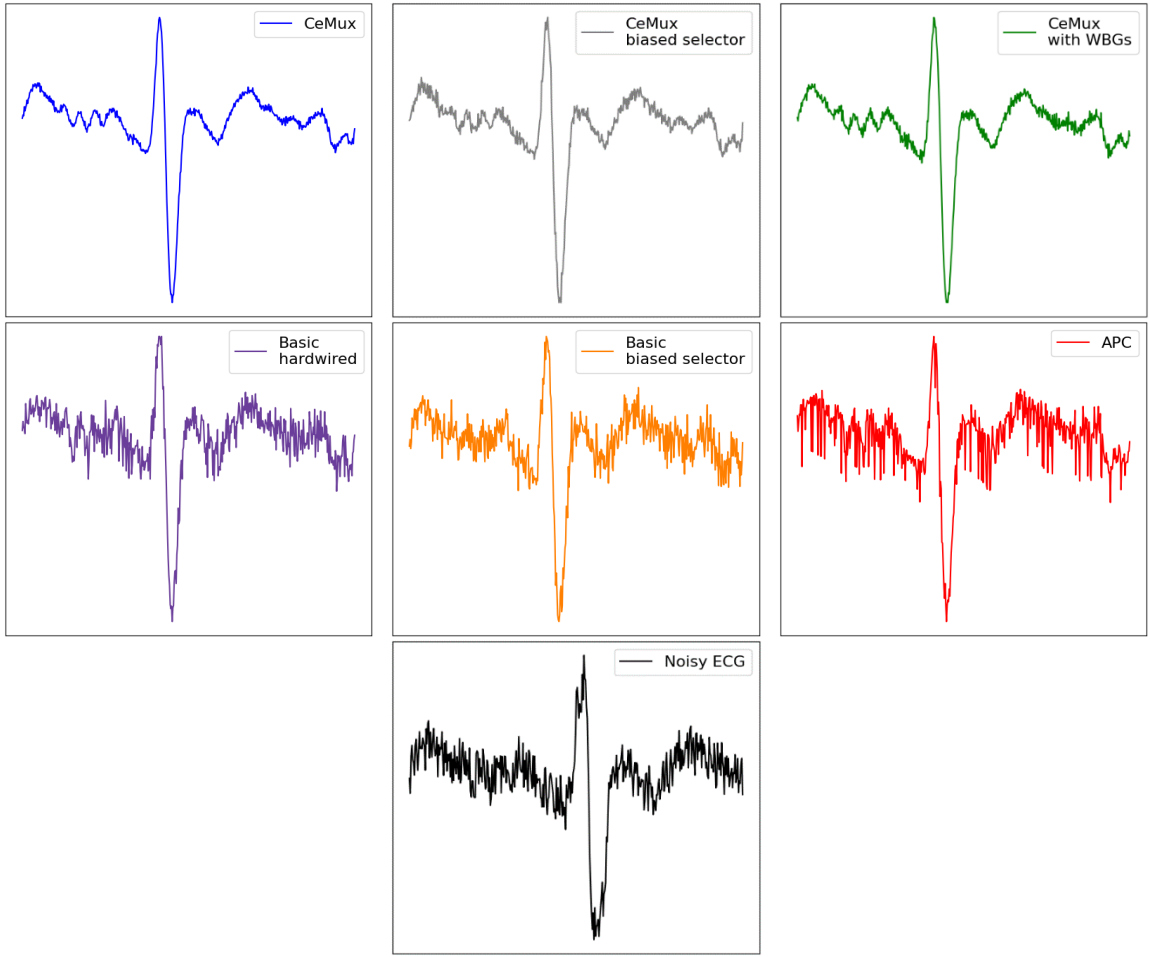

Figure 16: Noisy ECG waveform (bottom) filtered by various 10-bit precision SC designs; SN length is $2^{10}$.

## 6 CONCLUSION

As a closing, example, we compare the performance of all the SC filter designs considered here in the case of a noisy ECG signal filtered by an $M = 100$ tap filter with precision $n$ set to 10. As Fig. 16 shows, the CeMux-based filters produce the smoothest, most noise-free curves, another reflection of CeMux's superior accuracy. In general, we have seen that CeMux is the best SC design developed so far for large weighted-adder networks, in terms of both accuracy and accuracy-area tradeoffs. These properties result from two key design features: full data correlation and precise sampling, both of which exploit correlation in new ways to reduce errors in stochastic signals. CeMux can thus be considered a major step towards practical implementation of many-input, compact adders for a variety of SC applications.

**ACKNOWLEDGEMENTS**

This research was supported by the U.S. National Science Foundation under Grant CCF-2006704.

# APPENDICES

Here we derive analytic expressions for mux adder variance and various adder configurations using the Bernoulli and hypergeometric SN models. Expressions derived with the hypergeometric SN model match the simulated variance of circuits that employ LFSR SNGs but overestimate the variance of CeMux when a low discrepancy random number source is used.

In the following derivations, a bipolar SN's bits are defined to take values $\{-1,1\}$ rather than $\{0,1\}$ where $-1$ acts as logical 0. Consequently, an $N$-bit bipolar SN $\mathbf{X}$'s estimated value becomes $\frac{1}{N}\sum_{i=1}^{N} X_i$ and its expected bit value $\mathbb{E}[X_i]$ becomes $\mu_{X_i}$ both of which match the unipolar case. Ultimately, changing the definition of bipolar bits allows some of the following expressions to simultaneously apply to both unipolar and bipolar SNs. The analysis also assumes that the bits of all input SNs are identically distributed which is the case in both the Bernoulli and hypergeometric SN models. It is helpful to note that the expectation operation, $\mathbb{E}[\cdot]$ is linear.

**A1. Mux Variance Decomposition (**13**)**

Consider a mux tree with $M$ data input SNs, $\mathbf{X}_1\mathbf{X}_2...\mathbf{X}_M$ that have values $\mu_{X_1}, \mu_{X_2}, ..., \mu_{X_M}$ and length $N$. The mux tree has a select input $\mathbf{S}$ which is a stream of identically distributed random words. $\mathbf{S}$'s value determines which data input is selected during clock cycle $j$. Let $|\widetilde{w}_i|$ be the probability that $\mathbf{X}_i$ is sampled during any given clock cycle. Let the output of the mux tree be SN $\mathbf{Z} = [Z_1, Z_2, ..., Z_N]^T$ whose estimated value is $\hat{\mu}_Z = \frac{1}{N}\sum_{i=1}^{N} Z_i$. The variance of the mux tree output estimator is

$$\text{Var}(\hat{\mu}_Z) = \mathbb{E}[(\hat{\mu}_Z - \mathbb{E}[\hat{\mu}_Z])^2] \tag{16}$$

Let $C_i$ be a random variable representing the number of times that $\mathbf{X}_i$ is sampled by the mux tree. First, it can be shown that since the input SN bits are identically distributed, the output SN's value can expressed as follows.

$$\hat{\mu}_Z = \frac{1}{N}\sum_{i=1}^{N} Z_i = \frac{1}{N}\sum_{i=1}^{M}\sum_{j=1}^{C_i} X_{i,j} \tag{17}$$

where $X_{i,j}$ represents $\mathbf{X}_i$'s bit when it is sampled for the $j^{th}$ time rather than $\mathbf{X}_i$'s $j^{th}$ bit.

$$\mathbb{E}[\hat{\mu}_Z] = \frac{1}{N}\sum_{i=1}^{M} \mathbb{E}\left[\sum_{j=1}^{C_i} X_{i,j}\right] \tag{18}$$

Because $C_i$ is a random variable, $\mathbb{E}\left[\sum_{j=1}^{C_i} X_{i,j}\right]$ is a random sum of random variables which evaluates to $\mathbb{E}[C_i]\mathbb{E}[X_{i,j}]$. Further, since $C_i$ is the number of times $\mathbf{X}_i$ is sampled and $|\widetilde{w}_i|$ is the probability that $\mathbf{X}_i$ is sampled during any given clock cycle, we have $\mathbb{E}[C_i] = |\widetilde{w}_i|N$. Putting these notions together yields

$$\mathbb{E}\left[\sum_{j=1}^{C_i} X_{i,j}\right] = \mathbb{E}[C_i]\mathbb{E}[X_{i,j}] = |\widetilde{w}_i|N\mu_{X_i} \tag{19}$$

Thus, (18) becomes

$$\mathbb{E}[\hat{\mu}_Z] = \sum_{i=1}^{M} |\widetilde{w}_i|\mu_{X_i} \tag{20}$$

In other words, we have that the mux's expected output value is a weighted sum of its input values as expected. The output variance (16) can then be rewritten using (17) and (20)

$$\text{Var}(\hat{\mu}_Z) = \mathbb{E}\left[\left(\frac{1}{N}\sum_{i=1}^{M}\sum_{j=1}^{C_i} X_{i,j} - \sum_{i=1}^{M}|\widetilde{w}_i|\mu_{X_i}\right)^2\right] \quad (21)$$

Now define $\epsilon_i = \frac{1}{N}\sum_{j=1}^{C_i} X_{i,j} - |\widetilde{w}_i|\mu_{X_i}$. Then

$$\text{Var}(\hat{\mu}_Z) = \mathbb{E}\left[\left(\sum_{i=1}^{M}\epsilon_i\right)^2\right] = \mathbb{E}\left[\sum_{i=1}^{M}\sum_{j=1}^{M}\epsilon_i\epsilon_j\right] \quad (22)$$

$$\text{Var}(\hat{\mu}_Z) = \sum_{i=1}^{M}\mathbb{E}[\epsilon_i^2] + \sum_{i=1}^{M}\sum_{\substack{j=1\\j\neq i}}^{M}\mathbb{E}[\epsilon_i\epsilon_j] \quad (23)$$

First re-express $\mathbb{E}[\epsilon_i^2]$.

$$\mathbb{E}[\epsilon_i^2] = \mathbb{E}\left[\left(\frac{1}{N}\sum_{j=1}^{C_i} X_{i,j} - |\widetilde{w}_i|\mu_{X_i}\right)^2\right] \quad (24)$$

Noting (19) and the definition of variance we have

$$\mathbb{E}[\epsilon_i^2] = \frac{1}{N^2}\text{Var}\left(\sum_{j=1}^{C_i} X_{i,j}\right) \quad (25)$$

$\sum_{j=1}^{C_i} X_{i,j}$ is again a random sum of random variables. Since the $X_{i,j}$'s are identically distributed and independent of $C_i$, it can be shown that

$$\mathbb{E}[\epsilon_i^2] = \frac{1}{N^2}\left(\text{Var}\left(\sum_{j=1}^{\mathbb{E}[C_i]} X_{i,j}\right) + \text{Var}(C_i)\mathbb{E}[X_{i,j}X_{i,k}]\right) \quad (26)$$

Now re-express $\mathbb{E}[\epsilon_i\epsilon_j]$

$$\mathbb{E}[\epsilon_i\epsilon_j] = \mathbb{E}\left[\left(\frac{1}{N}\sum_{k=1}^{C_i} X_{i,k} - |\widetilde{w}_i|\mu_{X_i}\right)\left(\frac{1}{N}\sum_{l=1}^{C_j} X_{j,l} - |\widetilde{w}_j|\mu_{X_j}\right)\right] \quad (27)$$

Expanding yields

$$\mathbb{E}[\epsilon_i\epsilon_j] = \mathbb{E}\left[\frac{1}{N^2}\sum_{k=1}^{C_i} X_{i,k}\sum_{l=1}^{C_j} X_{j,l} - \frac{|\widetilde{w}_j|\mu_{X_j}}{N}\sum_{k=1}^{C_i} X_{i,k} - \frac{|\widetilde{w}_i|\mu_{X_i}}{N}\sum_{l=1}^{C_j} X_{j,l} + |\widetilde{w}_i|\mu_{X_i}|\widetilde{w}_j|\mu_{X_j}\right] \quad (28)$$

Using (19),

$$\mathbb{E}[\epsilon_i\epsilon_j] = \frac{1}{N^2}\mathbb{E}\left[\sum_{k=1}^{C_i}\sum_{l=1}^{C_j} X_{i,k}X_{j,l} - N^2|\widetilde{w}_i|\mu_{X_i}|\widetilde{w}_j|\mu_{X_j}\right] \quad (29)$$

Noting $\sum_{k=1}^{C_i} \sum_{l=1}^{C_i} X_{i,k} X_{j,l}$ is a random sum of random variables, $\mathbb{E}[C_i] = |\widetilde{w}_i| N$ and $\mathbb{E}[X_{i,j}] = \mu_{X_i}$

$$\mathbb{E}[\epsilon_i \epsilon_j] = \frac{1}{N^2} \left[ \mathbb{E}[C_i C_j] \mathbb{E}[X_{i,k} X_{j,l}] - \mathbb{E}[C_i] \mathbb{E}[C_j] \mathbb{E}[X_{i,k}] \mathbb{E}[X_{j,l}] \right] \tag{30}$$

Noting the definition of covariance for two random variables $A$ and $B$, $\text{Cov}(A, B) = \mathbb{E}[AB] - \mathbb{E}[A]\mathbb{E}[B]$

$$\mathbb{E}[\epsilon_i \epsilon_j] = \frac{1}{N^2} \left[ \text{Cov}(C_i, C_j) \mathbb{E}[X_{i,k} X_{j,l}] + \mathbb{E}[C_i] \mathbb{E}[C_j] \text{Cov}(X_{i,k}, X_{j,l}) \right] \tag{31}$$

Putting (23), (26) and (31) together yields

$$\text{Var}(\hat{\mu}_Z) = \frac{1}{N^2} \left[ \sum_{i=1}^{M} \text{Var}\left( \sum_{j=1}^{\mathbb{E}[C_i]} X_{i,j} \right) + \sum_{i=1}^{M} \text{Var}(C_i) \mathbb{E}[X_{i,j} X_{i,k}] + \sum_{i=1}^{M} \sum_{\substack{j=1 \\ j \neq i}}^{M} \left[ \text{Cov}(C_i, C_j) \mathbb{E}[X_{i,k} X_{j,l}] + \mathbb{E}[C_i] \mathbb{E}[C_j] \text{Cov}(X_{i,k}, X_{j,l}) \right] \right]$$

For a random variable $A$, $\text{Var}(A) = \text{Cov}(A, A)$, hence the second summation $\sum_{i=1}^{M} \text{Var}(C_i) \mathbb{E}[X_{i,j} X_{i,k}]$ and first term in the final double summation can be combined.

$$\text{Var}(\hat{\mu}_Z) = \frac{1}{N^2} \left[ \sum_{i=1}^{M} \text{Var}\left( \sum_{j=1}^{\mathbb{E}[C_i]} X_{i,j} \right) + \sum_{i=1}^{M} \sum_{j=1}^{M} \text{Cov}(C_i, C_j) \mathbb{E}[X_{i,k} X_{j,l}] + \sum_{i=1}^{M} \sum_{\substack{j=1 \\ j \neq i}}^{M} \mathbb{E}[C_i] \mathbb{E}[C_j] \text{Cov}(X_{i,k}, X_{j,l}) \right] \tag{32}$$

Define

$$\epsilon_{noise} = \frac{1}{N^2} \sum_{i=1}^{M} \text{Var}\left( \sum_{j=1}^{\mathbb{E}[C_i]} X_{i,j} \right) \tag{33}$$

$$\epsilon_{samp} = \frac{1}{N^2} \sum_{i=1}^{M} \sum_{j=1}^{M} \text{Cov}(C_i, C_j) \mathbb{E}[X_{i,k} X_{j,l}] \tag{34}$$

$$\epsilon_{corr} = \frac{1}{N^2} \sum_{i=1}^{M} \sum_{\substack{j=1 \\ j \neq i}}^{M} \mathbb{E}[C_i] \mathbb{E}[C_j] \text{Cov}(X_{i,k}, X_{j,l}) \tag{35}$$

where $C_i$ is the number of times $\mathbf{X}_i$ is sampled and $X_{i,k}$ is the $k$-th sampled bit of $\mathbf{X}_i$ (not the $k$-th bit of $\mathbf{X}_i$). This redefinition of $X_{i,k}$ is permitted because both the Bernoulli and hypergeometric SN models assume bits are identically distributed. Note that we defined that $\mathbb{P}(\mathbf{X}_i \text{ is sampled}) = |\widetilde{w}_i|$. Eqs (33–35) apply to both the unipolar SN and bipolar SN cases, but in the bipolar SN case, the bits take value $\{-1,1\}$ instead of $\{0,1\}$ where $-1$ acts as logical 0. Finally, if an XNOR array is used before the mux tree as in Fig. 4, (33–35) still apply, but $X_{i,j}$ is redefined to be $\text{sign}(w_i) X_{i,j}$ where $\text{sign}(w_i) = 1$ if $w_i \geq 0$ and $\text{sign}(w_i) = -1$ otherwise.

$\epsilon_{noise}$ only depends on the variance of the input SNs which is determined by the SN model (i.e., Bernoulli or hypergeometric). $\epsilon_{samp}$ depends mainly on the covariance of the number of times each input is sampled which is determined by the sampling method (noisy or precise). $\epsilon_{corr}$ is a function of the covariance between sampled bits of two input SNs which depends on the SN model and on the SCC between input SNs. In all, we have

$$\text{Var}(\hat{\mu}_{Z_i}) = \epsilon_{noise} + \epsilon_{samp} + \epsilon_{corr} \tag{36}$$

## A2. Expressions for Variance Components

In Table 3, we list expressions for $\text{Var}(\hat{\mu}_{Z_i})$ when the mux tree has bipolar inputs and when an XNOR array is used before the mux tree. To derive such expressions, $\epsilon_{noise}$, $\epsilon_{samp}$, and $\epsilon_{corr}$ are re-expressed according to which SN model (Bernoulli or hypergeometric), sampling method (noisy or precise) and input correlation level (SCC = 0 or SCC = +1) is used. Then $\epsilon_{noise}$, $\epsilon_{samp}$ and $\epsilon_{corr}$ are summed together. For precise sampling, we assume that $|\widetilde{w}_i|N$ is an integer, which is always the case if $N$ is a power of 2 and a hardwired mux tree is used. Input correlation of SCC = 0 means the pairwise SCC between all mux tree inputs is 0 and input correlation SCC = +1 means the pairwise SCC between all mux tree inputs is +1 (as is the case when full correlation is achieved). Note that these derived equations were experimentally validated by simulating the stochastic circuits in which they correspond to.

Finally, for the SCC +1 case, it is helpful to define the order statistics [30] of the mux tree input. Let $s_i = \text{sign}(w_i)$ and let $A = \{s_0\mu_{X_0}, s_1\mu_{X_1}, \dots, s_M\mu_{X_M}\}$ be the values of the SN inputs to the mux tree. Then $s_{(i)}\mu_{X_{(i)}}$ is defined to be the $i$-th order statistic of $A$. For instance, $s_{(0)}\mu_{X_{(0)}}$ is the minimum element in $A$, $s_{(M)}\mu_{X_{(M)}}$ is the maximum element in $A$ and, in general, $s_{(i)}\mu_{X_{(i)}}$ is the $i$-th largest element in $A$.

Of the six equations presented in Table 3, only the first corresponding to the Bernoulli model with noisy sampling has appeared in SC literature before [7][23]. Inspecting the equations in Table 3 reveals that switching from Bernoulli to hypergeometric input SNs decreases variance except in atypical cases like when all SNs have value +1, all have value −1 in the bipolar case or all have value 0 in the unipolar case. In those cases, variance stays the same when switching from Bernoulli to hypergeometric SNs. Likewise, both switching from noisy to precise sampling and switching from input SCC level 0 to level +1 decreases variance except in atypical cases where variance stays the same. Thus, according to the hypergeometric and Bernoulli SN models, using precise sampling and achieving full correlation are always beneficial.

Interestingly, when full correlation (SCC level +1) is achieved for hypergeometric SNs, the variance is a function of the *difference* between input SN values rather than a function of the input SN values themselves as in other cases. This implies that if the input SN values are similar, the variance is smaller. Indeed, in the example of Fig. 8b, the input SNs all have the same value and variance in that case is zero.

Table 3: Derived Variances for Bipolar Mux Adders

| Input SN Model | Sampling Method | Input SCC Level | Derived Output Variance |
|---|---|---|---|
| Bernoulli | Noisy | Any | $\dfrac{1 - \left(\sum_{i=1}^{M} \widetilde{w}_i \mu_{X_i}\right)^2}{N}$ |
| Bernoulli | Precise | Any | $\dfrac{1 - \sum_{i=1}^{M} |\widetilde{w}_i| \mu_{X_i}^2}{N}$ |
| Hypergeometric | Noisy | 0 | $\dfrac{1 - \left(\sum_{i=1}^{M} \widetilde{w}_i \mu_{X_i}\right)^2 - \sum_{i=1}^{M} \widetilde{w}_i^2 (1 - \mu_{X_i}^2)}{N}$ |
| Hypergeometric | Noisy | 1 | $\dfrac{\sum_{i=1}^{M} \sum_{j=1}^{i-1} |\widetilde{w}_i||\widetilde{w}_j|\left(s_{(j)}\mu_{X_{(j)}} - s_{(i)}\mu_{X_{(i)}}\right)}{N}$ |
| Hypergeometric | Precise | 0 | $\dfrac{\sum_{i=1}^{M} |\widetilde{w}_i|(1 - |\widetilde{w}_i|)(1 - \mu_{X_i}^2)}{N - 1}$ |
| Hypergeometric | Precise | 1 | $\dfrac{\sum_{i=1}^{M} \sum_{j=1}^{i-1} |\widetilde{w}_i||\widetilde{w}_j|\left(s_{(j)}\mu_{X_{(j)}} - s_{(i)}\mu_{X_{(i)}}\right)\left(1 - \left(s_{(j)}\mu_{X_{(j)}} - s_{(i)}\mu_{X_{(i)}}\right)\right)}{N - 1}$ |